\documentclass[nonacm,manuscript]{acmart}
\AtBeginDocument{%
  }

\usepackage{booktabs}
\usepackage{graphicx}
\usepackage{subcaption}
\usepackage{bm}
\usepackage{listings}   
\usepackage{enumitem}   
\usepackage{xcolor}     
\usepackage{tabularx}   
\usepackage{booktabs}   
\usepackage{geometry}   
\lstdefinelanguage{json}{
    frame=leftline,
    basicstyle=\ttfamily\small,
    stepnumber=1,
    showstringspaces=false,
    breaklines=true,
    frame=single,
    morestring=[b]",
    stringstyle=\color{black},
}
\begin{document}

\title[TaskSense: Cognitive Chain Modeling and Difficulty Estimation...]{TaskSense: Cognitive Chain Modeling and Difficulty Estimation of GUI Tasks} 

\author{Yiwen Yin}
\email{yinyw21@mails.tsinghua.edu.cn}
\orcid{0009-0002-0620-7902}
\affiliation{%
  \institution{Tsinghua University}
  \country{China}
}

\author{Zhian Hu}
\email{hza25@uw.edu}
\affiliation{%
  \institution{University of Washington}
  \city{Seattle}
  \state{Washington}
  \country{USA}
}

\author{Xiaoxi Xu}
\affiliation{%
  \institution{Cornell University}
  \city{Ithaca}
  \state{New York}
  \country{USA}
}

\author{Chun Yu}
\email{chunyu@tsinghua.edu.cn}
\orcid{0000-0003-2591-7993}
\affiliation{%
  \institution{Tsinghua University}
  \city{Beijing}
  \country{China}
}

\author{Xintong Wu}
\affiliation{%
  \institution{Tsinghua University}
  \city{Beijing}
  \country{China}
}

\author{Wenyu Fan}
\email{wfan0971@uni.sydney.edu.au}
\affiliation{%
  \institution{University of Sydney}
  \city{Sydney}
  \country{Australia}
}

\author{Yuanchun Shi}
\email{shiyc@tsinghua.edu.cn}
\affiliation{%
  \institution{Tsinghua University}
  \city{Beijing}
  \country{China}
}

\renewcommand{\shortauthors}{Yin. et al.}

\begin{abstract}


Measuring GUI task difficulty is crucial for user behavior analysis and agent capability evaluation. Yet, existing benchmarks typically quantify difficulty based on motor actions (e.g., step counts), overlooking the cognitive demands underlying task completion. In this work, we propose \textbf{Cognitive Chain}, a novel framework that models task difficulty from a cognitive perspective. A cognitive chain decomposes the cognitive processes preceding a motor action into a sequence of cognitive steps (e.g., finding, deciding, computing), each with a difficulty index grounded in information theories. We develop an LLM-based method to automatically extract cognitive chains from task execution traces. Validation with linear regression shows that our estimated cognitive difficulty correlates well with user completion time (step-level $R^{2}$=0.46 after annotation). Assessment of state-of-the-art GUI agents shows reduced success on cognitively demanding tasks, revealing capability gaps and Human-AI consistency patterns. We conclude by discussing potential applications in agent training, capability assessment, and human-agent delegation optimization.

\end{abstract}


\begin{teaserfigure}
  \includegraphics[width=\textwidth]{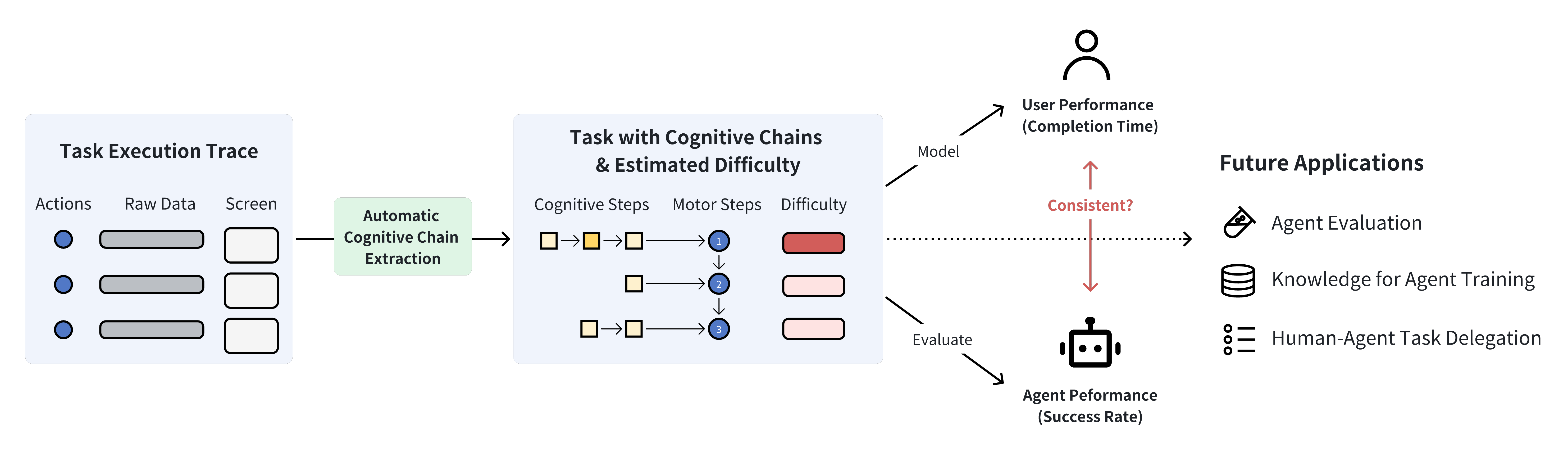}
  \caption{We propose Cognitive Chain, a novel framework for estimating cognitive difficulty for GUI tasks by modeling the cognitive processes that precedes each motor action, along with an automatic cognitive chain extraction method. Evaluations on both user and agent performance demonstrate the expressiveness of our model, and reveal patterns of Human-AI consistency regarding cognitive capabilities.}
  \Description{This is the description of the teaser.}
  \label{fig:teaser}
\end{teaserfigure}


\maketitle
\section{Introduction}

Modeling and estimating the difficulty of tasks on Graphical User Interface (GUI) has long been fundamental to human–computer interaction research\cite{card1980klm, card1983goms}. Traditional theories and models assess task difficulty to better understand user behaviour and predict user performance on interfaces, thereby informing interface design, evaluation, and user training\cite{fitts1954information, hick1952rate, card1980klm, card1983goms, byrne2001act-r/pm, card1986mhp, li2018predicting-menu, do2021simulation-point}. 
More recently, with the emergence of GUI agents--systems that automate GUI tasks powered by Large Language Models (LLMs) and Vision-Language Models (VLMs)\cite{qin2025ui-tars, anthropic2024computeruse, openai2025cua}--task difficulty has become a central dimension for evaluating agent performance. Modern benchmarks often categorize tasks based on estimated difficulty to enable systematic comparisons of agent capabilities across diverse scenarios\cite{deng2023mind2web, he2024webvoyager, pan2024webcanvas, mialon2023gaia, xie2024osworld}.

However, existing agent benchmarks typically define task difficulty solely by the number of steps, overlooking the \textbf{cognitive demands} inherent in GUI tasks.
A GUI task often encompasses not only motor actions (e.g., clicking or typing), but also a range of cognitive processes involved in interaction which demand mental effort and reasoning capabilities.
For example, when users create a calendar event using information from an email, they must first extract and retain event details (e.g., subject and time), switch to the calendar, recall that information, locate the correct time slot, and finally compose an appropriate event entry (see Figure~\ref{fig:cogchain}).  
By capturing different cognitive activities involved--such as information extraction, element finding, content creation, and decision making--and their respective demands, we can more precisely estimate task difficulty and better reveal the cognitive limitations of current GUI agents.

Although several theories and models provide more refined techniques for estimating task difficulty, they still fall short in addressing cognitive demands, as they primarily focus on motor actions\cite{fitts1954information, card1980klm} or narrowly scoped tasks such as menu selection\cite{li2018predicting-menu, do2021simulation-point}.
Moreover, cognitive-oriented modeling frameworks tend to offer general-purpose abstractions that either lack a close examination of GUI-specific interactions\cite{byrne2001act-r/pm}, or rely on oversimplified definitions of cognitive processes\cite{yin2025taskmind}.
We argue that GUI task difficulty should be revisited from a cognitive perspective, enabling a more comprehensive understanding of both user performance and agent capabilities.



In this paper, we introduce \textbf{Cognitive Chain}, a novel framework for estimating GUI task difficulty by modeling the cognitive processes underlying GUI interactions. 
A cognitive chain refers to the cognitive activities that occur before each motor action, consisting of a sequence of cognitive steps such as element identification, decision making and content creation.
We introduce a taxonomy of 8 cognitive types commonly involved in GUI interactions, and define type-specific methods to compute the difficulty index of each cognitive step based on context-dependent factors, grounded in information-theoretic principles.
A linear model is employed to quantify task difficulty as the sum of cognitive step difficulties across all cognitive chains. 
To support scalable application, we also develop an LLM-based method for automatically extracting cognitive chains from task execution traces. 
Our model captures both the structure and difficulty of the cognitive processes preceding GUI actions, enabling cognitively grounded estimation of task difficulty for both user performance analysis and agent capability evaluation.

Using a curated dataset of 18 GUI tasks, we conducted two experiments to evaluate our model's expressiveness and generalizability from both user and agent perspectives.
First, we collected user interaction traces with extracted cognitive chains, and empirically validated our linear model by fitting type-specific base difficulty parameters on actual step completion times, along with cross-validation. We evaluated both our automatic extraction method and the cognitive chain model using raw outputs and human-annotated data, respectively. Comparing to a baseline that assumes a uniform average step time, results show that cognitive chain-derived difficulty correlate well with both step-level and task-level durations, achieving regression R-squared scores of 0.28 (raw) and 0.46 (annotated), with prediction RMSE values of 46.1\% (raw) and 37.4\% (annotated). 
We also observed the need for richer contextual cues for GUI cognition understanding, such as user attention indicators.
Second, we evaluated four state-of-the-art GUI agents on the same task set and observed consistent performance drops on steps with higher cognitive demand, particularly in specific types. These results highlight both the homogeneity and heterogeneity of cognitive mechanisms between humans and agents, and reveal that current agents still struggle with cognitively intensive subtasks.
We conclude by discussing potential applications of our cognitive chain model and automatic extraction method for future directions such as agent training, cognitive benchmarking, and human-agent task delegation.

Overall, our contributions are:
\begin{itemize}
    \item The cognitive chain model, a novel framework for estimating GUI task difficulty by modeling the cognitive processes underlying GUI interactions. 
    \item An LLM-driven method for automatically extracting cognitive chains from task execution traces, with a practical implementation.
    \item A comprehensive evaluation of the cognitive chain model using human completion times and GUI agent performance, demonstrating its expressiveness in capturing task difficulty, while revealing consistency patterns in cognitive capabilities between human and agents.
\end{itemize}
\section{Related Work}

\subsection{Semantic Understanding of GUI and GUI Actions} 


Understanding the semantics of recorded UI screens and GUI actions is fundamental to GUI task modeling and difficulty estimation. 
Prior work on GUI understanding has taken two main directions. Some approaches leverage web DOM or accessibility data to capture the semantics of GUIs\cite{barman2016ringer, little2007koala, yin2025taskmind}, while others\cite{yeh2009sikuli, li2021screen2vec, leiva2020enrico, bai2021uibert, wang2021screen2words, li2020widget} rely purely on visual information to interpret screens and elements. For example, models such as Screen2vec\cite{li2021screen2vec} and UIBert\cite{bai2021uibert} embed UI screenshots into semantic representations that enable downstream applications, and recent GUI agents\cite{openai2025cua, anthropic2024computeruse, qin2025ui-tars} have advanced by training multimodal large language models to understand screenshots during execution.

The understanding of GUI actions can be viewed on two levels. First level of work focuses on semantics of user interactions, typically by mapping recorded user events to structured representations of actions defined by researchers, using rules or machine learning models\cite{yin2025taskmind, he2021actionbert}. Other works\cite{zhang2024designwatch} attempt to infer action semantics from raw screen recordings without explicit event logs. The second level goes beyond action semantics to uncover users’ high-level intentions. For instance, Programming by Demonstration systems\cite{li2018appinite, li2019pumice} interpret the intention of task actions to support task automation and generalization.

In this paper, we leverages the multimodal capabilities of LLMs to jointly interpret UI screenshots and accessibility data. This allows us to capture both the semantics and intentions of GUI actions, providing a foundation for extracting cognitive chain from user interaction traces.

\subsection{GUI Task Modeling and Difficulty Estimation} 




A number of classical task theories and models provide foundational perspectives for understanding and analyzing GUI-based tasks. 
Among these, GOMS\cite{card1983goms} decomposes tasks into a hierarchy of goals, subgoals, and the operators (actions) required to achieve them. 
The Keystroke-Level Model (KLM)\cite{card1980klm} simplifies GOMS and focuses on low-level keystroke and mouse-click operations.
The Model Human Processor (MHP)\cite{card1986mhp}, ACT-R/PM\cite{byrne2001act-r/pm}, and the OODA loop\cite{bazin2005boyd-ooda} emphasize cognitive and decision-making processes during task execution. 
Norman’s Seven Stages of Action\cite{norman2013design} offers a conceptual framework that describes the goal formation, execution and evaluation phases.
In addition, several works\cite{paterno1997concurtasktrees, georgievski2014htn} enable the modeling of specific types of tasks including concurrent and hierarchical tasks.
By modeling user interactions with GUI, these models serve as a systematic foundation for assessing task difficulty and guiding interface design.

Many of these models have been extended to assess task difficulty, with task completion time often serving as an empirical measure.
For instance, Fitts’ Law\cite{fitts1954information} and Hick’s Law\cite{hick1952rate} provide quantitative measures of manual task difficulty, focusing on motor control and decision-making time, respectively. 
Other studies\cite{hoffmann1997concurrent, thomas1974combined-manual-decision} further explore the integration of motor and decision tasks, analyzing how their concurrent execution affects task difficulty.
To account for cognitive activities, KLM\cite{card1980klm} estimates task execution time by decomposing a task into a sequence of low-level operators, each associated with a time cost. Cognitive activity is treated as a distinct operator, typically modeling a fixed time taken for mental preparation before a physical action.
MHP\cite{card1986mhp} decomposes a processing cycle into \textbf{perceptual, cognitive, and motor} subsystems, each with empirically derived average times. 
The ACT-R/PM model\cite{byrne2001act-r/pm} further distinguishes cognitive activity between procedural and declarative memory processes, and provides a general framework for estimating their time costs.
Recent methods combine modern techniques such as deep learning to predict user performance on interfaces\cite{li2018predicting-menu, do2021simulation-point}.
However, these models focus primarily on perceptual and motor performance, lacking fine-grained categorizations and difficulty estimates for the diverse cognitive types involved in GUI-based interactions.




In practice, existing GUI agents benchmarks\cite{deng2023mind2web, pan2024webcanvas, he2024webvoyager, xie2024osworld, mialon2023gaia} divide task difficulty levels based primarily on the number of actions required. 
Dataset such as OSWorld\cite{xie2024osworld} and GAIA\cite{mialon2023gaia} provide a basic analysis of the capabilities needed for different tasks, but these are mostly limited to perceptual or execution-level abilities (e.g. file reading, web browsing), or domain-specific software knowledge, rather than cognitive capabilities.
While these metrics provide a basic understanding of task difficulty and capability requirements, they fail to capture the nuanced cognitive challenges inherent in GUI interactions. 
This limitation highlights the need for more sophisticated models and evaluation metrics that can account for the cognitive demands involved in completing GUI-based tasks.




\subsection{Cognitive Chain in GUI Tasks} 




Chain-of-Thought (CoT)\cite{wei2022cot} is both a concept and a practice designed to improve the ability of LLMs by prompting them to generate intermediate reasoning steps before the final answer. By mimicking human cognitive processes, CoT has been proved effective, and is applied across various LLM tasks. 
For example, cognitive chain is used to enhance model performance and explainability in psychological stress detection\cite{wang2024cognitionchain-stress} and social situation reasoning tasks\cite{park2025cognitivecot-social}.
The ReAct framework\cite{yao2023react} demonstrates that integrating reasoning steps into models capable of taking actions can improve performance on downstream tasks. 
Furthermore, CodeI/O\cite{li2025codeio} showed that the benefits of using reasoning patterns from code generation as training data can generalize effectively to other tasks.

In this work, we use the concept of ``cognitive chain'' to describe the human reasoning process that precedes each action during task execution. 
Studies of human cognitive activity have a rich foundation, including classic cognitive theories such as Bloom’s Taxonomy\cite{bloom1964taxonomy}, Marzano \& Kendall’s Framework\cite{marzano2006newtaxonomy}, and Multiple Intelligence Theory\cite{bornstein1986frames}. 
Understanding and modeling human cognition during GUI-based tasks holds value not only for cognitive science but also for improving task efficiency and designing better interactive systems.
Previous work has explored cognition modeling in GUI tasks such as menu selection\cite{kangasraasio2017cogmodels-ABC} and repetitive workflows\cite{li2018appinite, yin2025taskmind}. 
However, these efforts typically focus on specific task domains, or rely on simplified cognitive process definitions. In contrast, our work proposes a taxonomy of cognitive step types for general GUI tasks and provides difficulty estimates for each type, contributing to a deeper understanding of cognitive chains in GUI interactions.


\section{Formative Study and Cognitive Chain}
We first define our cognitive process modeling and difficulty estimation problem. We then conduct a think-aloud study to uncover cognitive patterns users exhibit while performing GUI tasks. Building on these insights, we propose the concept of \textbf{cognitive chain} to model the cognitive processes and the associated difficulty involved in GUI tasks.

\subsection{Problem Formulation}

A GUI task is typically composed of a sequence of motor steps, such as clicking buttons or entering text. However, beneath these explicit actions lies an implicit layer of cognitive processes that guide user behavior.
The problem we aim to address is: \textbf{Given the task environment (i.e. user interface) and a known method for completing the task (i.e., a sequence of steps), how can we infer the cognitive processes that precede each motor step, and estimate their associated difficulty?}

Here, ``method'' refers to a series of operators (action steps) used to achieve a task goal, as defined in GOMS models\cite{card1983goms}, where a task goal may be accomplished through multiple alternative methods. In our problem definition, we assume that the method is already known.
Furthermore, to support downstream applications that may require real-time inference of cognitive demands, we limit the context to only the previously executed steps, the current step, and optionally a short future window of upcoming steps. The window length can be set to any value, including zero.
Note that these steps do not need to have been actually executed--our model can also be applied to a system (e.g., a task agent) that can predict upcoming steps for a task goal. The step prediction is beyond the scope of our work.

Formally and specifically, for each motor step, the input includes:
\begin{itemize}
    \item[-] All prior steps,
    \item[-] A short window of upcoming steps (optional),
    \item[-] The current GUI screen,
    \item[-] The current motor step.
\end{itemize}
The output consists of:
\begin{itemize}
    \item[-] The cognitive processes that occur prior to the motor step,
    \item[-] The estimated difficulty of the cognitive processes.
\end{itemize}

\subsection{Formative Study}
\label{section:formative_study}



We recruited 5 participants (3 female, 2 male), aged between 19 and 26 ($\sigma$=3.0), from a university campus to complete computer-based tasks in a lab setting under the observation of a researcher.
Each participant was asked to perform common tasks across three major themes: planning a trip, conducting research, and creating presentation slides.
Participants selected specific tasks they were familiar with under each theme. 
During task execution, participants were required to articulate what they were thinking while performing each action step. 
We recorded both their GUI interactions and verbal utterances.
The study lasted about 1.5 hour for each participant, and they were compensated at CNY 100 per hour.\footnote{The study protocol was reviewed and approved by the university ethics review board. The compensation was consistent with the average earning of workers in the community where the study took place.}

In total, we collected 7.5 hours of screen and audio recordings. We then analyzed the cognitive processes expressed during these sessions, focusing on two core questions: (1) What types of cognitive activities are involved in GUI tasks, and what structural patterns do they exhibit in relation to motor steps? (2) How can the difficulty of these cognitive activities be estimated?
We derived several key insights that guided the design of our cognitive modeling framework:
\begin{itemize}
    \item \textbf{Cognitive processes can be decomposed into multiple steps.} They are typically linear, and occur either before or after a motor step. Occasionally, however, cognitive and motor steps occur simultaneously--for example, while searching for specific content on the screen, users tended to make clicks to facilitate the process.
    \item \textbf{Cognitive steps can be categorized into distinct types}, including goal formulation, information or element search, memory recall, decision-making, etc. 
    \item \textbf{Difficult cognitive steps require longer thinking time, and the factors that contribute to the difficulty differ by cognitive type.} These factors can often be inferred from the semantic intent of the motor step, and the information available on the current interface. For instance, the difficulty of searching or making a decision is often related to the number of candidate options, whereas reorienting the next sub-goal during execution may depend on the amount of progress made and the complexity of the remaining goals.
\end{itemize}
Building on these findings, we recognized a set of common cognitive steps and proposed the concept of ``cognitive chain'' to describe the cognitive processes and their difficulty during GUI interactions.
We detail the cognitive chain model in the following sections.

\subsection{Cognitive Chain}
\label{section:cogchain}

\begin{figure*}[htbp]
    \centering
    \includegraphics[width=1\linewidth]{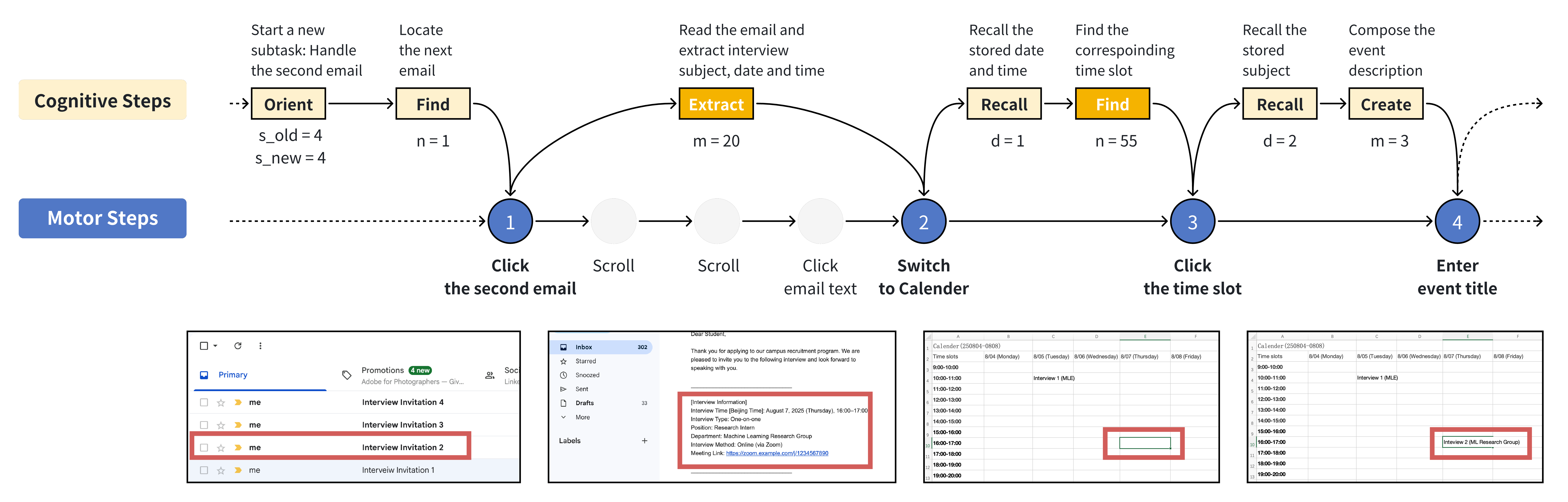}
    \caption{An example of cognitive chains in a task (T15 in our task set), where the user creates calendar events using interview invitation information from emails.}
    \label{fig:cogchain}
\end{figure*}

\begin{table*}[htbp]
\centering
\caption{Cognitive types and their corresponding difficulty index estimation methods.}
\label{table:cogsteps}
\renewcommand{\arraystretch}{1}
\begin{tabular}{@{}p{1.2cm}p{4.1cm}p{3.4cm}p{5.4cm}@{}}
\toprule
Type & Definition  & Example  & Difficulty Index Function $I(factors)$ 
\\ \midrule

\textbf{Orient}         
& Process of orienting towards the next (sub)goal and planning steps to achieve it           
& \begin{minipage}{3.6cm}
    Orient the next line to fill
  \end{minipage} 
& $\bm{log(s_{old}) + log(s_{new})}$, where $s_{old}$ denotes the number of completed steps, $s_{new}$ denotes the number of upcoming steps 
\\
\addlinespace[1pt] 
\hline
\addlinespace[1pt] 

\textbf{Find}
& Process of locating specific elements within the interface. 
& \begin{minipage}{3.6cm}
    Find the font size button 
  \end{minipage}
& $\bm{log(n+1)}$, where 
$n$ denotes the number of candidate elements (If the element has been located before, $n$ degrades to 1)  
\\
\addlinespace[1pt] 
\hline
\addlinespace[1pt] 

\textbf{Extract}
& Process of identifying specific information from text or multimodal content
& \begin{minipage}{3.6cm}
    Get location from document
  \end{minipage}
& $\bm{log(m+1)}$, where 
$m$ denotes the number of candidate information chunks, which is determined by the amount of information relative to the target information to be extracted. 
\\
\addlinespace[1pt] 
\hline
\addlinespace[1pt] 


\textbf{Recall}         
& Process of retrieving information from memory
& \begin{minipage}{3.6cm}
    Recall the previous copy
  \end{minipage}
& $\bm{1-e^{-d/t}}$, where 
$d$ denotes the steps between the last memory storage and the current step, $t$ is a constant representing memory decay rate 
\\
\addlinespace[1pt] 
\hline
\addlinespace[1pt] 

\textbf{Decide}         
& Process of making a choice, whether between multiple options or without explicit options. 
& \begin{minipage}{3.6cm}
    Decide among hotels 
  \end{minipage}
& $\begin{cases}
\bm{log(n+1)} & \text{for explicit options} \\
\bm{c} & \text{for implicit decisions}
\end{cases}$, where 
$n$ denotes the number of candidate options, $c\in[0,1]$ denotes the estimated size of implicit decision space 
\\
\addlinespace[1pt] 
\hline
\addlinespace[1pt] 

\textbf{Compute}        
& Process of performing calculations, comparison or logical reasoning.                           
& \begin{minipage}{3.6cm}
    Compute the maximum 
  \end{minipage}
& $\bm{c}$, where 
$c\in[0,1]$ denotes the estimated computational complexity 
\\
\addlinespace[1pt] 
\hline
\addlinespace[1pt] 

\textbf{Create}  
& Process of generating new content
& \begin{minipage}{3.6cm}
    Write comments 
  \end{minipage}
& $\bm{log(m+1)}$, where 
$m$ denotes the number of information chunks to create 
\\
\addlinespace[1pt] 
\hline
\addlinespace[1pt] 

\textbf{Verify}         
& Process of checking the correctness of motor actions
& \begin{minipage}{3.6cm}
    Verify the created plot
  \end{minipage}
& $\bm{log(m+1)}$, where      
$m$ denotes the number of information chunks to verify 
\\ \bottomrule

\end{tabular}%
\end{table*}

A cognitive chain is an ordered sequence of cognitive steps occur before each motor steps during GUI task execution. 
A Cognitive step represents a unit of cognitive activity, and we categorize all cognitive steps into 8 types, as shown in Table~\ref{table:cogsteps}.
This categorization is informed by insights from our formative study as well as prior works about cognitive taxonomy\cite{bloom1964taxonomy, yin2025taskmind}. 
A GUI task can thus be viewed as a series of \textbf{cognitive chain-motor step} cycles, where each motor action is guided by a cognitive process.
Figure~\ref{fig:cogchain} illustrates an example of cognitive chains in the task \textit{} (T15 from our dataset in Section~\ref{section:dataset}).


Note that while most cognitive steps precede motor actions, some may occur concurrently, as discussed in Section~\ref{section:formative_study}. Therefore, in such cases, a single cognitive step may span multiple motor steps. 
For example, as shown in Figure~\ref{fig:cogchain}, actions between motor step 1 and 2 represent a continuous cognitive process of information extraction, while scrolling and clicking serve as auxiliary actions. These actions are covered by a single cognitive step labeled ``Extract''.
Conversely, some motor steps may not be preceded by any cognitive steps. This often happens when a high-level action (e.g., multiple selection) is decomposed into smaller sub steps, and the user has completed the cognitive processing before the first sub-step. The remaining sub-steps are then carried out with minimal deliberation.



\subsection{Difficulty Estimation}



We identify a set of difficulty factors for each cognitive type, along with a method to compute the \textbf{difficulty index} $\bm{I}$ for each cognitive step, as summarized in Table~\ref{table:cogsteps}.
The estimation of the difficulty index is informed by classical cognitive theories. 
For example, the difficulty of Decision steps with explicit options is inspired by Hick’s Law\cite{hick1952rate}, which states that decision time increases logarithmically with the number of choices. 
For other types involving information processing such as Find and Extract, information theory\cite{shannon1948mathematical} is also applied to calculate the entropy of the information involved as the difficulty index. 
The difficulty of Recall steps is guided by memory decay models, particularly Ebbinghaus' Forgetting Curve\cite{ebbinghaus2013memory} and the Model Human Processor\cite{card1986mhp}.
Note that for all cases involving information quantity $m$, we use the number of chunks as the unit--i.e., meaningful blocks of information\cite{thalmann2019chunk}. This serves as a relative measure. For instance, in Extract steps, $m$ is estimated as the ratio between the amount of the total information and the target information to be selected.
Furthermore, for Find steps, we distinguish between known-location and unknown-location scenarios. If the user has a rough idea of its location (e.g., from earlier actions), the difficulty index is reduced to 1, reflecting a minimal search cost, as exemplified by Step 1 in Figure~\ref{fig:cogchain} where the location of the second email is known because the user has already clicked on the first one.



Each cognitive type has a constant base difficulty $\bm{K^{Type}}$ (to be empirically fitted in Section~\ref{section:regression}), and the difficulty of each cognitive step is defined as:
\begin{equation}
    D^{\text{CogStep}} = K^{\text{Type}} \cdot I^{\text{CogStep}}
    \label{eq:cogstep}
\end{equation}
Overall, we model the difficulty of each task step as the linear sum of its cognitive steps and the motor step. The cognitive chain captures thinking processes, while the motor component reflects motor preparation and execution.
The total task difficulty is modeled as the sum of all step difficulties. This linear assumption follows prior work\cite{hoffmann1997concurrent}, as our GUI tasks are self-paced with no external time pressure, where cognitive and motor steps are independent and do not influence each other.
Formally, the task and step difficulties are defined as:
\begin{equation}
 D^{\text{Task}} = \sum_{i=1}^{n} D^{\text{Step}}_{i}
\label{eq:task}
\end{equation}
\begin{equation}
 D^{\text{Step}}_{i} = D^{\text{CogChain}}_{i} + D^{\text{Motor}}_{i} = \sum_{j=1}^{m_i} D_{i,j}^{\text{CogStep}} + D^{\text{Motor}}_{i}
\label{eq:step}
\end{equation}
where $n$ denotes the number of motor steps in a task, $m_i$ denotes the number of cognitive steps in the $i$-th cognitive chain, and $D^{\text{CogStep}}_{i,j}$ is the difficulty of the $j$-th cognitive step within that chain.


\section{Cognitive Chain Extraction Method}
\begin{figure*}[htbp]
    \centering
    \includegraphics[width=\linewidth]{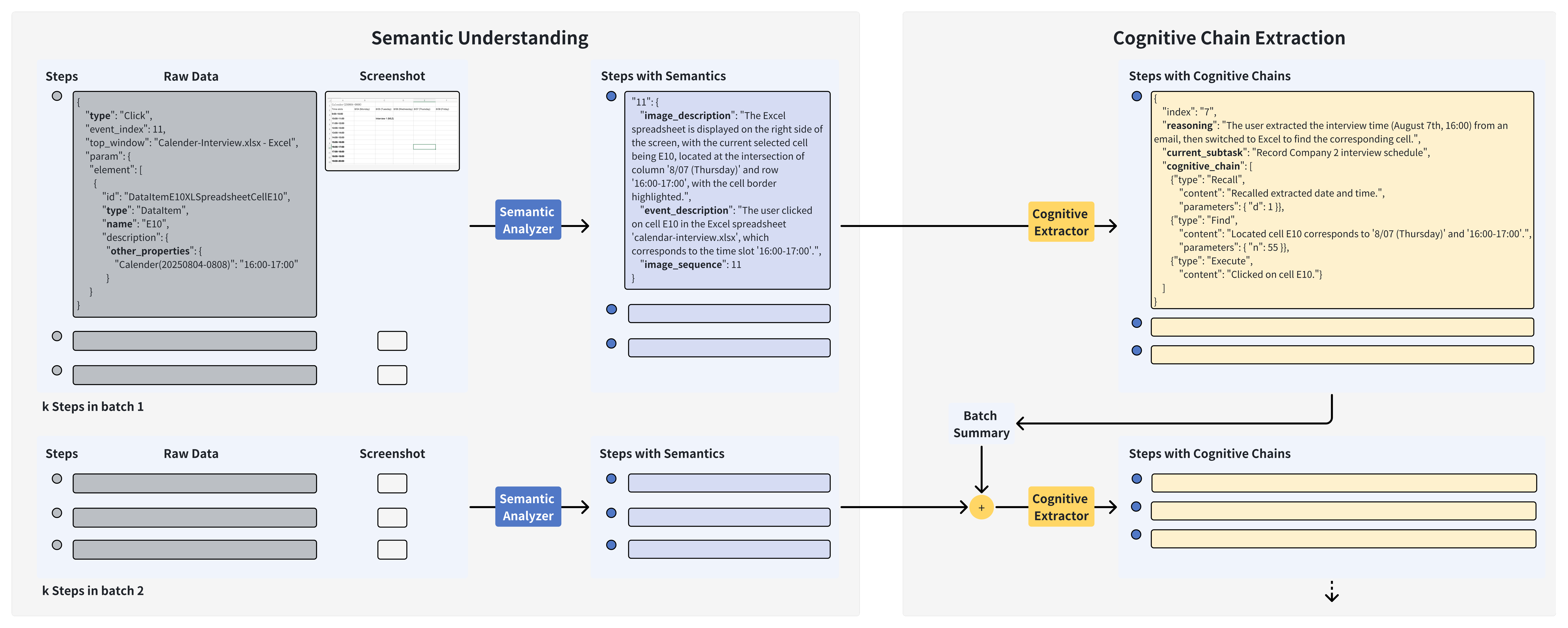}
    \caption{Workflow of our cognitive chain extraction method.}
    \label{fig:method}
\end{figure*}

In this section, we detail our method of extracting cognitive chains with difficulty factors from user interaction traces. 
Our extraction system consists of two major steps: semantic understanding and cognitive chain extraction. The workflow is shown in Figure \ref{fig:method}.

\subsection{Semantic Understanding}
\label{section:semantic_understanding}
We first use an LLM-based semantic analyzer to interpret the semantics of each user motor action.
The input includes basic metadata for each keyboard and mouse event, along with associated screen captures.
The output is natural language descriptions of semantics for motor actions. We use GPT-4o as our base model. 
The prompt provided to the model contains detailed instructions, which are listed in Appendix~\ref{appendix:prompt_semantic_analyzer}.


Due to the context length limitations of LLMs, we divide the user actions into batches and feed them into the LLM incrementally, then assemble the outputs into a complete task representation. The batch size $k$ is set to 10. This batch-based design has two key advantages: (1) It improves the output accuracy by limiting semantic drift over long sequences. (2) It allows for flexible batch sizing, which enables real-time analysis of user actions for further downstream applications such as real-time AI agent support (see Section~\ref{section:applications}).

\subsection{Cognitive Chain Extraction}
Base on the semantic description of each motor action, we apply an LLM-based cognitive chain extractor to identify the cognitive steps and its estimated difficulty (See Section \ref{section:cogchain}). 

Similar to the semantic understanding phase, we process steps in batches ($k$=10). In addition to the current step batch, we also provide the LLM with a summary of the previous batch to ensure continuity and preserve global task context.
Thus, the input to the cognitive extractor includes: the semantics of motor steps in the current batch, corresponding screenshots, and a summary of previous steps. The output contains: the intention and the cognitive chain consisting of sequential cognitive steps for each motor action, and estimated difficulty factors for each cognitive step, according to the cognitive type (e.g. n for Find, and c for Compute, see Table~\ref{table:cogsteps}). 
Figure~\ref{fig:method} shows an example of the input and output JSON format. 
We use Gemini-2.5-pro as our base model.
The prompt instructs the model to (1) analyze steps sequentially, (2) identify the step intention, (3) segment actions into distinct subtasks, and (4) extract the cognitive chain with estimated difficulty factors.
We then parse and assemble the resulting JSONs into the final structured output. The full prompt is provided in Appendix~\ref{appendix:prompt_cogchain_extractor}.
\section{Dataset}
\label{section:dataset}

We constructed a set of 18 representative GUI tasks to further assessment. These tasks were carefully selected to cover a wide range of cognitive types and difficulty levels, enabling systematic evaluation of both human users and GUI agents. 
To reflect common real-world workflows, we drew insights from both natural user interaction logs and existing benchmark datasets.



\textbf{Real-world Task Collection.} 
We collected real computer usage data from 33 participants (aged 20-53, $\mu$=26.8, $\sigma$=6.8) with diverse occupations such as students, teachers, counselors and freelancers.
Participants installed our logging software, which recorded screenshots, keyboard and mouse activities when launched. 
They were instructed to use their computers as usual over a one-week period, with the freedom to start or stop recording at any time. 
All participants were informed about potential privacy implications, provided informed consent before the collection, and were compensated at CNY 30 per hour for recording.\footnote{The study protocol was reviewed and approved by the university ethics review board. The compensation was consistent with the average earning of workers in the community where the study took place.}
In total, we collected 69.9 hours of interaction logs. We segmented these into 212 distinct tasks with the help of an LLM, and identified those suitable for automation--i.e., tasks with clear goals and structured workflows.

\textbf{Existing Datasets.} We also incorporated examples from commonly used datasets for GUI agent assessment, including OSWorld\cite{xie2024osworld} and Online-Mind2Web\cite{pan2024webcanvas}, to broaden task diversity and ensure alignment with current benchmarks.

\textbf{Final Task Set.} We synthesized a set of 18 tasks that simulate realistic workflows, such as transferring data between spreadsheets, collecting product reviews, and designing posters (see Appendix~\ref{appendix:taskset}).\footnote{All data including user traces and extracted cognitive chains will be released after publication.}
To facilitate large-scale trace collection, tasks were designed around scenario that common users are likely to be familiar with.
All tasks are based on widely used office applications, including word processors, spreadsheets, presentation tools, web browsers, and file managers.
We ensured that the task set covers all cognitive types and different difficulty levels defined in Section~\ref{section:cogchain}, with at least six tasks for each type.

\section{Empirical Validation on User Performance}




We collected user interaction traces on the task set in a controlled laboratory setting, and used actual step completion times to validate our cognitive chain model and extraction method. Our evaluation addresses two core research questions:
\begin{itemize}
    \item (RQ1) How accurately can our method automatically extract cognitive chains from user interaction traces?
    \item (RQ2) To what extent does our cognitive chain model reflect actual task difficulty, as evidenced by the cognitive time for each step and the entire task?
\end{itemize}
We first applied our automatic extraction method to the collected traces, then had human annotators review and refine the resulting cognitive chains.
We fit the model-estimated difficulties (see Section~\ref{section:cogchain}) to users' actual cognitive times, and used the base difficulty ($\bm{K_{type}}$) fitted from all but one participant to predict the held-out user's cognitive times via cross-validation. 
We applied this procedure to both the raw and annotated data to answer the two research questions.

\subsection{Trace Collection Procedure}
we collected interaction traces from different users on the same computer in a controlled lab study, ensuring consistent task familiarity and interface complexity while minimize external confounds such as cognitive fatigue and environmental distractions. 
We recruited 6 participants (3 males and 3 females) from the campus community, with different majors and an average age of 23.3 ($\sigma$=1.2). 
We limited the number of participants to six due to the high cost of manually annotating step-level data, and the large number of tasks and steps per task ensured sufficient data for statistical analysis.
All participants were required to be familiar with the Windows system, reporting an average daily usage of 6.1 hours ($\sigma$=3.5). 
The study began with a brief introduction session, followed by a free-use session to help participants get familiar with the lab computer environment and essential software. 
Participants were then given a paper detailing 18 tasks with their goals and required methods. 
Before the formal recording of each task, they were required to fully understand and memorize the goal and method. 
They also performed a practice version before each task with different data to ensure familiarity with the task procedure, while preventing exposure to the actual data that could bias their cognitive performance. 
Participants were allowed to take breaks between tasks as needed.
They were compensated at CNY 100 per hour for recording.\footnote{The study protocol was reviewed and approved by the university ethics review board. The compensation was consistent with the average earning of workers in the community where the study took place.}

During formal recordings, we used our recording software to capture screens with accessibility data, and timestamps for each keyboard and mouse event. 
A rule-based method was applied to group related events into higher-level continuous steps, such as combining consecutive keyboard events into a single "TextInput" step, with start and end timestamps.
In total, we collected 108 task traces (6 participants × 18 tasks).

\textbf{Data Pre-processing.}
For each task trace, we first used the semantic analyzer to interpret the semantics of each motor step, followed by manual review and refinement. The refinement process involved: (1) correcting misinterpreted semantics, and (2) removing clearly erroneous or irrelevant steps. 
We also calculated the actual completion time $T^{\text{step}}_i$ for $i$-th motor step, defined as the elapsed time from the end of the previous step to the start of the action (i.e., the first mouse or keyboard event), based on recorded timestamps:
\begin{equation}
    T^{\text{step}}_i = t^{\text{start}}_{i} - t^{\text{end}}_{i-1} %
    \label{eq:steptime}
\end{equation}
Note that we have excluded the execution durations for continuous actions such as text input or dragging, by separately recording their start and end timestamps. The resulting step time reflects both cognitive processing and potential motor preparation.
The final dataset consists of interaction traces with accurate semantic labels and aligned step-level timing, providing the foundation for further cognitive chain extraction and analysis.




\subsection{Evaluation Methodology}
We first applied our cognitive chain extractor to all interaction traces. Then, 4 researchers who were fully familiar with the definition of cognitive chains manually reviewed and refined the raw outputs. Note that annotators were blind to the actual step times during the annotation process. 

Following annotation, we used the actual step times $\bm{T^{\text{Step}}_i}$ as the dependent variable and fit the base difficulty parameters $\bm{K^{\text{Type}}}$ using linear regression to evaluate our difficulty estimation model. Recall that in our model, step difficulty is a linear sum of weighted cognitive step difficulties (see Equation~\ref{eq:cogstep} and \ref{eq:step}):
\begin{equation}
    T^{\text{Step}}_i = \sum_{j=1}^{m_i} K_{i,j}^{\text{Type}} \cdot I_{i,j}^{\text{CogStep}} + D^{\text{Motor}}_i
    \label{eq:regression}
\end{equation}
The type and difficulty index $\bm{I_{i,j}^{\text{CogStep}}}$ of each cognitive step are known inputs obtained from the cognitive chains.
We apply this regression both to the raw outputs (to assess the accuracy of our automatic extraction method, RQ1),and to the annotated cognitive chains (to validate our cognitive chain model's expressiveness, RQ2).

Beyond fitting on the full data, we also conducted leave-one-subject-out (LOSO) cross-validation to evaluate the predictive performance of our model. For each fold, the model was trained on $N-1$ participants and tested on the held-out participant. 
Considering that users exhibit different basic response and motor execution times, we reserved 5 tasks (28\% of the validation set) per participant as calibration data. From these, we fitted a linear mapping: 
\begin{equation}
 T^{\text{observed}} = \alpha \cdot T^{\text{predicted}} + \beta
\label{eq:calib}
\end{equation}
where $\alpha$ and $\beta$ denote participant-specific slope and intercept, which were then applied to adjust the final predictions.

As a baseline, we compare against the model that assumes all motor steps take equal time, regardless of cognitive steps and difficulty (step-mean baseline). For fairness, we also applied the same per-user linear calibration to the baseline predictions during cross-validation.
We also include a second baseline--the cognitive chain model \textbf{without difficulty estimation}, which assigns a uniform difficulty index of 1 to all cognitive steps, assuming that steps of the same type share a constant base difficulty. This allows us to isolate and evaluate the expressiveness of our difficulty estimation.

\subsection{Results}
\label{section:regression}

\begin{table*}[htbp]
    \centering
    \caption{Regression and prediction results. Note that the minor difference between raw and annotated data of step-mean baseline is due to corrections of a few actions occurred concurrently with cognitive steps.}
    \label{tab:time_results}
    \begin{tabular}{@{}lcccccc@{}}
    \toprule
    \multicolumn{1}{c}{}           & \multicolumn{2}{c}{Regression $R^2$ (Step)} & \multicolumn{2}{c}{Regression $R^2$ (Task)} & \multicolumn{2}{c}{LOSO Prediction RMSE (Task)} \\
    \multicolumn{1}{c}{}           & Raw                    & Annotated                  & Raw                    & Annotated                  & Raw                     & Annotated                 \\ \midrule
    Step-mean Baseline             & 0.00                   & 0.00                       & 0.36                   & 0.40                       & 59.2\%                  & 58.2\%                    \\
    Cognitive Chain w/o Difficulty & 0.25                   & 0.36                       & 0.57                   & 0.63                       & 46.7\%                  & 42.8\%                    \\
    \textbf{Cognitive Chain}                & \textbf{0.28}                   & \textbf{0.46}                       & \textbf{0.61}                   & \textbf{0.69}                       & \textbf{46.1\%}                  & \textbf{37.4\%}                    \\ \bottomrule
    \end{tabular}%
\end{table*}

\begin{figure*}[htbp]
    \centering
    \begin{subfigure}[t]{0.48\linewidth}
        \centering
        \includegraphics[width=\linewidth]{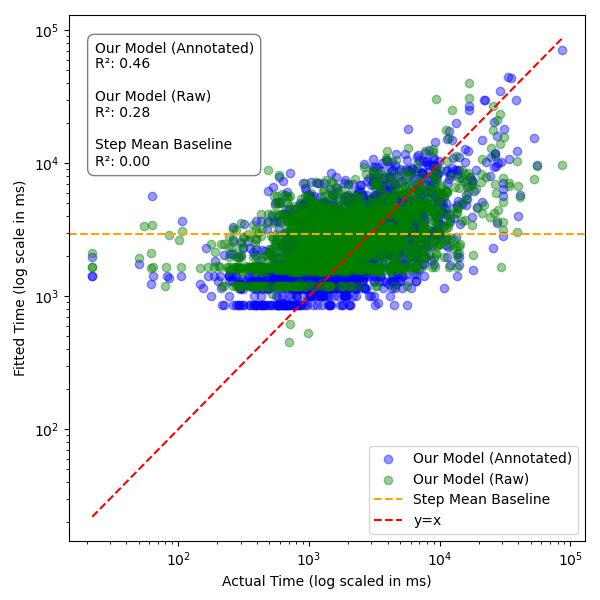}
        \caption{Actual and Fitted Time (Step-level)}
        \label{fig:fit-step}
    \end{subfigure}
    \hfill
    \begin{subfigure}[t]{0.48\linewidth}
        \centering
        \includegraphics[width=\linewidth]{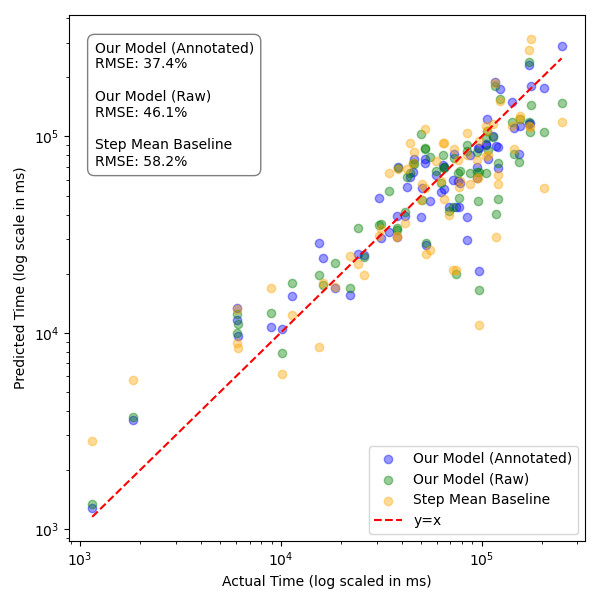}
        \caption{Actual and Predicted Time (Task-level)}
        \label{fig:predict-task}
    \end{subfigure}
    \caption{Visualized time regression and prediction results.}
    \label{fig:regression}
\end{figure*}

\begin{table}[htbp]
\caption{Fitted base difficulties of cognitive types.}
\label{tab:basetime}
\centering
    \begin{tabular}{@{}lc@{}}
    \toprule
    Cognitive Type   & Base Difficulty $K^{\text{Type}}$  \\ \midrule
    Orient           & 4.7 ms                 \\
    Find             & 563.2 ms               \\
    Extract          & 1415.9 ms              \\
    Recall           & 446.6 ms               \\
    Decide (explicit) & 742.0 ms               \\
    Decide (implicit) & 1506.4 ms              \\
    Compute          & 5120.1 ms              \\
    Create           & 1422.1 ms              \\
    Verify           & 778.1 ms               \\
    Intercept        & 859.1 ms               \\ \bottomrule
    \end{tabular}%
\end{table}


\subsubsection{Regression Results}
We first performed linear regression on a total of 2360 step samples from 108 task traces (21.9 steps per task in average, $\sigma$=14.1). The resulting $R^{2}$ values were 0.46 for the raw data and 0.28 for the annotated data, as shown in Table~\ref{tab:time_results}. 
Prior work on modeling human behaviour and cognition typically reports $R^{2}$ values ranging from 0.3 to 0.6\cite{awada2023predicting, peters2023context-prediction}. Given the diversity of tasks and the natural variability in users' step-level cognitive timing, our model achieves a reasonably good fit and demonstrates meaningful explanatory power for step-level time.
Since our step-mean baseline aligns with the null model used in $R^{2}$ computation (i.e., assumes the mean step duration), any positive $R^2$ indicates improvement over the step-mean baseline.
In addition, comparing to the cognitive chain without difficulty estimation, our results reveals that incorporating difficulty indices significantly improves task modeling accuracy.
Figure~\ref{fig:fit-step} visualizes the fitted and actual step times across all tasks, where logarithmic scales are used, as errors tend to scale proportionally with step or task duration\cite{card1980klm}.

We also summed fitted step times in a task and compared them against task-level actual times. The $R^{2}$ values across 108 tasks was 0.61 before and 0.69 after annotation. Comparing to 0.36 and 0.40 of a baseline model that assumes constant step time, our cognitive chain model yields significantly improved performance. Note that the minor difference between raw and annotated data in the step-mean baseline results is due to corrections of a few actions occurred concurrently with cognitive steps, as mentioned in Section~\ref{section:cogchain}.

The fitted base difficulty parameters (in milliseconds) are reported in Table~\ref{tab:basetime}. The intercept can be interpreted as the motor preparation time independent of cognitive processing. Using a difficulty index of 1 as reference, we find that Creative, Compute, and Decide steps tend to have longer cognitive time, while Orient and Recall steps are associated with shorter durations. These observations align with patterns noted in our formative study.

\subsubsection{Cross-Validation Results}
For the LOSO cross-validation, we summed the predicted step times to compute task durations and evaluated the prediction across all folds (78 tasks in total). At the task-level, the root-mean-squared error (RMSE) was 37.4\% (annotated) and 46.1\% (raw) of the average task time, which is acceptable for behavioral prediction tasks and shows substantial improvement over the baseline (58.2\% and 59.2\%). The comparison with the cognitive chain without difficulty estimation also highlights the effectiveness of our difficulty index. Final results were averaged across participants and reported in Table~\ref{tab:time_results}, with prediction performance visualized in Figure~\ref{fig:predict-task}.

\subsubsection{Error Analysis}


While our model demonstrates reasonable explanatory power on the annotated data, its performance on raw data remains limited. This highlights a key limitation that our automatic extraction method sometimes fail to fully capture the underlying cognitive chains involved in user tasks. By comparing the raw and annotated outputs, we observed several issues regarding RQ1:
\begin{enumerate}
    \item[1.] \textbf{Inaccurate Estimation of Task Difficulty.} The system often struggle to estimate certain difficulty factors. For example, when users engage in prolonged scrolling to inspect large amount of content in T9, LLMs may fail to account for the amount of all the information.
    \item[2.] \textbf{Lack of Multimodal Cues for Cognitive Chain Extraction.} Our system lacks access to rich multimodal data that reflect users’ cognitive state and task intent. For instance, when users select sample text and spends time inspect font styles in the toolbar, this Extract process may be missed without explicit motor actions. This suggests that attention-revealing signals, such as eye-tracking data, could serve as valuable complements to infer cognitive chains.
\end{enumerate}


Additionally, there are intrinsically unobservable cognitive processes that challenge even human annotators, resulting in the limited performance of our model (RQ2). These processes are latent and transient, often influenced by momentary confusion, fatigue, or attention shifts--factors not easily inferred from behavior alone. 
For example, cognitive steps such as Verify and Recall may be ambiguous, and occur intermittently with actions. In T1, it is often unclear whether users verify information before or after taking motor actions, and the cognitive rhythm varies greatly across users.
Decide is also difficult to pinpoint--some users unconsciously select a text before deciding what to do with it; others make deliberate decisions first. 
Moreover, some cognitive steps may occur concurrently with motor actions (e.g., thinking while typing), which our current linear cognitive model does not accommodate.


\section{Assessment of Agent Performance}







To assess whether our difficulty modeling framework can generalize beyond human users, we evaluated four state-of-the-art GUI agents on the same set of tasks, including Anthropic Claude 4 (virtual Linux)\cite{anthropic2024computeruse}, Anthropic Claude 4 (real Windows deployment),  UI-TARS\cite{qin2025ui-tars}, and Fellou (on Windows)\cite{fellou2025ai}. Our research questions are:
\begin{itemize}
    \item (RQ3) How do current agents perform on different types and levels of cognitive steps?
    \item (RQ4) Can our difficulty estimation framework be applied to assess agent capabilities?
\end{itemize}
Since agent action traces sometimes deviate from human demonstrations due to redundant actions or incorrect attempts, we standardized the evaluation by identifying the essential motor steps with cognitive chains required to complete each task. These steps were distilled from user traces and define the minimum cognitive path necessary for successful task completion, which was inspired by prior work\cite{pan2024webcanvas}.

To accurately capture agent performance on different cognitive steps, we conducted manual evaluations of action success, and, in failure cases, identified the specific cognitive step where the error originated.
Steps that were never attempted due to early failures, or that failed as a consequence of earlier cognitive errors, were excluded from analysis. Steps omitted due to the agent’s self-initiated task termination were marked as failures.
For tasks involving generative content (e.g., draft a title for a poster in T4), we also assessed whether the generated result met user expectations.

\begin{figure*}[htbp]
    \centering
    \includegraphics[width=0.9\linewidth]{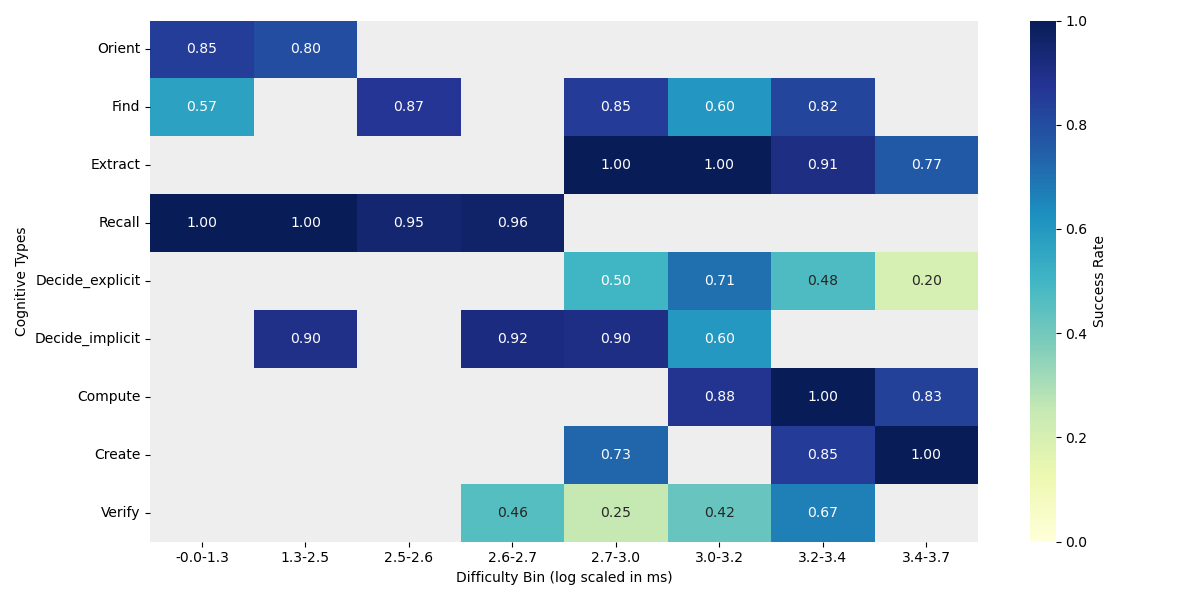}
    \caption{Success rate of four agents across different cognitive types and difficulty bins (equal-frequency).}
    \label{fig:agent}
\end{figure*}

Figure~\ref{fig:agent} shows the success rate of these agents across different cognitive types and difficulty levels. The x-axis represents the estimated cognitive difficulty, calculated using the base difficulty parameters derived in Section~\ref{section:regression}, and discretized into equal-frequency bins. Empty cells indicate that a particular cognitive type did not occur at that difficulty level. We summarize our key observations as follows:

\textbf{Current Agents Struggle with Certain Cognitive Demands (RQ3).} In summary:
\begin{itemize}
    \item Agents consistently underperform on Verify steps, often failing to correctly interpret GUI feedback to determine whether their actions had the intended effect, or satisfied the task goal. For example, in T1, the agent used a VLOOKUP function to move data but failed to notice the consequent error indicated by an “N/A” value.
    \item Orient failures typically occur after multiple steps, when agents lose track of task context or fail to plan effective recovery strategies after an error. 
    \item Some cognitive steps, such as Create and Decide, rely on \textbf{user-specific context} that agents cannot autonomously infer. For example, performance on hotel selection in T10 often reveals limitations, while writing reviews in T12 typically requires preferences from the user to produce satisfactory results. These failures are often due to the agent's inability to access user-specific preferences or contextual information necessary for informed decision-making.
\end{itemize}

\textbf{Alignment and Divergence Between Human and Agent Cognitive Patterns (RQ4).}
For cognitive types such as Orient, Decide, and Extract, agent success rates negatively correlate with estimated cognitive difficulty, suggesting that our human-centered cognitive difficulty model generalizes well for these categories. This indicates a degree of alignment in perceived difficulty between humans and agents.
Meanwhile, step types such as Find, Compute, and Verify show weaker alignment: agents may perform well even on steps that are cognitively demanding for humans, or exhibit uniformly average performance regardless of difficulty. For example, agents often struggled with Find steps involving UI element identification, while rarely fail with Compute steps. This indicates fundamental differences in cognitive patterns and execution strategies, reflecting cognitive \textbf{heterogeneity} between humans and agents.

\section{Discussion of Envisioned Applications} 
\label{section:applications}
\subsection{Agents' Capability Assessment} 



Our cognitive chain model provides a multi-dimensional evaluation of cognitive demands in GUI tasks, offering metrics that capture diverse abilities such as element search, decision-making, and content creation. This contributes a richer and more fine-grained framework for assessing GUI task difficulty for both humans and agents, effectively bridging traditional theories of task difficulty estimation with the emerging evaluation needs of frontier AI agents.
These multi-dimensional metrics can inform agent benchmarking and optimization. They not only suggest concrete optimization targets, but also guide AI development strategies: for instance, training with capability-specific data, designing multi-agent architectures that mirror human cognitive processes, and dynamically switching between small and expert models based on cognitive demands.
We also believe that similarly fine-grained cognitive demand metrics are essential in other task areas, such as deep research\cite{huang2025deep-research} and creativity\cite{zhang2023creative}. Our work on GUI tasks provides a reference point for developing tailored evaluation systems in other domains.

\subsection{Task Knowledge for Agent Improvement} 

The cognitive chains extracted by our method offer a deeper understanding of GUI tasks beyond manual-level interactions, capturing the underlying thinking processes during task completion.
We envision that these cognitive chains 
can serve as valuable training data for GUI-based or other task-completion agents, improving agent success rates by providing chain-of-thought data for intermediate steps and sub-goals. 
Additionally, inspired by recent work like ReAct\cite{yao2023react} and CODEI/O\cite{li2025codeio}, which demonstrate the benefits of reasoning data for downstream and even cross-application tasks, we expect that cognitive data from GUI tasks can similarly enhance broader machine learning applications.
Moreover, by capturing how users perceive UI feedback, process information, and take actions, cognitive chains could also serve as external knowledge to improve agents capabilities--for example, faster and more accurate element finding, greater decision-making, and effective recovery from errors during task execution.


\subsection{Human-Agent Task Delegation Optimization} 


Beyond providing training data, our method can also support more effective Human-AI collaboration for task agents by informing appropriate task delegation strategies.
By modeling the cognitive demands and difficulty of each step in a task plan, \textbf{an agent can determine whether a step should be automated or delegated to the user}. Steps with high cognitive demands (e.g., difficult decision-making), or require user-specific input (e.g. content creation based on user preference), may be better suited for user control or partial user involvement, while low-difficulty steps can be automated by the agent. Such adaptive task allocation enhances overall efficiency and cooperation between humans and AI systems.


Furthermore, our extraction method can be implemented as a \textbf{real-time user interaction understanding system} (see Section \ref{section:semantic_understanding}), enabling proactive AI that provides timely and context-aware assistance by observing user interactions and inferring their underlying cognitive state.
For example, when a user is filling a complex form and repeatedly searching for reference information, the agent could infer user's thinking process, estimate the cognitive demands, and proactively pre-fill relevant fields or even take over the task to reduce user effort.




\section{Limitations and Future Work}



Our model and method still face several limitations that requires further improvement.
First, understanding users’ cognitive steps based solely on their manual actions remains challenging. Important cues such as mouse and eye movement trajectories are not currently incorporated, yet they are crucial for accurately inferring users’ task intent and cognitive activities. 
Second, unintended and erroneous operations significantly affect task intention inference. Detecting such operations is inherently difficult, especially when they superficially resemble purposeful behavior.
In addition, our cognitive difficulty estimation model is relatively coarse. It assumes linear effects and does not yet account for potential interactions between motor operations and cognitive activities. More sophisticated models are needed to capture these complex dynamics in future work. 
\section{Conclusion}
In this paper, we introduced cognitive chain, a framework for user cognitive process modeling and difficulty estimation of GUI tasks. We implemented an LLM-based method to extract cognitive chains from GUI interaction traces. Through comprehensive evaluation using human completion time and agent success rates, we showed that cognitive chains offer deeper insights into task difficulty for both humans and AI agents. 
By combining symbolic theories of task difficulty with modern LLM-based task understanding, our approach not only provides multi-dimensional assessment metrics for GUI agents but also paves the way for more effective human–AI collaboration.



\bibliographystyle{ACM-Reference-Format}
\bibliography{main-bib}

\appendix
\clearpage
\onecolumn
\section{Task Set}
\label{appendix:taskset}

\begin{table*}[htbp]
\renewcommand{\arraystretch}{1}
    \centering
    \caption{Task list with descriptions and high-demand cognitive types.}
    \label{tab:task_list_translated}
    \begin{tabular}{@{}p{2.8cm}p{8cm}p{3.5cm}@{}}
        \toprule
        \textbf{Task ID} & \textbf{Task Description} & \textbf{High-demand Cognitive Types} \\
        \midrule
        T1-vlookup & Copy names and salaries from one table and paste them into another with a shuffled name order, ensuring correct matching. & Orient, Find, Extract, Recall \\
        \addlinespace
        T2-addplot & Based on the current data, insert a "scatter with smooth lines and markers" chart in Excel. & Find, Decide \\
        \addlinespace
        T3-period & Given the data, calculate the first column divided by the second, and highlight the maximum result value. & Decide, Compute \\
        \addlinespace
        T4-poster & Extract lecture content, time, and location from a Word document, paste it into a PPT using "keep source formatting", and create an engaging title. & Extract, Create, Verify \\
        \addlinespace
        T5-teacherinfo & Read a professor's encyclopedia profile and extract information: name, current institution, research field, contact information. & Extract \\
        \addlinespace
        T6-password & Following webpage instructions, navigate to a cloud drive link and enter the extraction code. & Recall, Find \\
        \addlinespace
        T7-formatdoc & Memorize "modify file format" steps and perform formatting adjustments sequentially in a document editor without looking back. & Orient, Recall, Find \\
        \addlinespace
        T8-removetmp & In a folder with various file types, delete only the ".tmp" files as instructed. & Decide \\
        \addlinespace
        T9-comment & Filter and select 5 positive product reviews that include descriptions of the product's flavor. & Decide, Find \\
        \addlinespace
        T10-roomdecide & Select and book a hotel that is inexpensive and well-located. & Decide \\
        \addlinespace
        T11-weekend & Highlight the weekends in a column of dates by changing the font color to red. & Compute \\
        \addlinespace
        T12-review & For the current document, provide ratings and written feedback across three dimensions. & Decide, Extract, Create \\
        \addlinespace
        T13-dataclean & Given a data table with messy formatting and missing items, standardize it. & Decide, Compute \\
        \addlinespace
        T14-emailreply & Read an email, extract key information, and write a reply by referencing a calendar. & Extract, Create \\
        \addlinespace
        T15-schedule & Read a series of four interview invitation emails and save them to the calendar. & Extract, Recall, Find \\
        \addlinespace
        T16-reimburse & Given two tables (actual expenses and reimbursement requests), find and mark entries where the amount or purpose does not match. & Decide, Compute, Extract \\
        \addlinespace
        T17-exampleformat & Adjust the font and font size of your document based on a sample file. & Orient, Extract, Recall \\
        \addlinespace
        T18-recruit & Write a new experiment recruitment notice by referencing a sample. & Decide, Create \\
        \bottomrule
    \end{tabular}
\end{table*}

\clearpage
\section{The Prompt of Semantic Analyzer}
\label{appendix:prompt_semantic_analyzer}

\textit{system}

\begin{quote}
\sloppy

\textbf{Role and Core Principles}

You are a top-tier expert in computer screen content recognition and user operation interpretation. Your core competencies are:
\begin{enumerate}
    \item \textbf{Log-First Analysis}: Always start with structured operational logs as the primary basis for analysis.
    \item \textbf{Visual Corroboration}: Use screenshots to provide visual context and detailed evidence for the operational logs.
    \item \textbf{Contextual Logic Chain}: Weave isolated operational events into a complete task workflow through contextual logic.
    \item \textbf{Universal UI Reasoning}: \textbf{When encountering unknown software, infer its functionality based on interface layout (e.g., navigation pane, main content area, properties panel) and element characteristics (e.g., icons, text).}
\end{enumerate}

\vspace{\baselineskip}

\textbf{Task Goal}

Your task is to receive a series of \texttt{event-image pairs} and generate a JSON analysis report for each event. The report must strictly follow a \textbf{one-to-one} principle: N input events must correspond to N JSON outputs.

\vspace{\baselineskip}

\textbf{Analysis Process: The Three-Step Method}

\begin{enumerate}
    \item \textbf{Decode the Log}: From the \texttt{event} record, extract \textbf{"Who (user) -> Where (application/window) -> How (action type) -> Operated What (UI element) -> With What Specifics (text/parameters)"}.
    \item \textbf{Focus on Visuals}:
        \begin{itemize}
            \item \textbf{View the image with the "operated object" in mind} to supplement its visual details and state.
            \item \textbf{Prioritize describing the visual focus}: \textbf{Describe the cursor position, highlighted/selected areas, and any newly appeared menus/windows first. The image description should follow a "holistic-focus-detail" hierarchy.}
        \end{itemize}
    \item \textbf{Construct the Narrative}:
        \begin{itemize}
            \item \textbf{Use the operational log as the skeleton}, fleshing it out with details from the image and contextual logic.
            \item \textbf{Adhere to the "Atomicity" principle}: \textbf{Clearly describe the "subject-verb-object" structure of the operation and quantify its parameters (e.g., the specific text entered, the exact format value modified).}
            \item \textbf{Link Related Facts}: \textbf{When describing sequential actions like copy-paste, you must explicitly state the content of the data being operated on by tracing back to preceding events.}
        \end{itemize}
\end{enumerate}

\vspace{\baselineskip}

\textbf{Specific Application Analysis Strategies}

\begin{itemize}
    \item \textbf{Spreadsheets/Excel}:
        \begin{enumerate}
            \item \textbf{Global Structure Memory}: Upon first encountering a table, \textbf{fully record all visible row and column headers} in the "image\_description" to establish a "row/column -> name" mapping memory.
            \item \textbf{Cross-Screenshot Inference}: In subsequent analyses, even if headers are scrolled out of view, you \textbf{must use the memorized mapping to accurately identify the full identity of the operated cell} in the "event\_description" (e.g., "User selected the 'Test Subject Fee' cell (I10) for 'Zhao Liu'").
        \end{enumerate}
    \item \textbf{PPT/Documents/Webpages}:
        \begin{enumerate}
            \item \textbf{Locate Core Elements}: Identify the core objects of the operation, such as text boxes, images, buttons, or input fields, from the operational record.
            \item \textbf{Extract Key Information}: In the "image\_description," describe the content, format, attributes, and relative position of the element on the page in detail.
        \end{enumerate}
\end{itemize}

\vspace{\baselineskip}

\textbf{Output Requirements}

\textbf{Important: Strictly Adhere to the Following:}
\begin{enumerate}
    \item \textbf{N input events must result in N JSON objects, one-to-one, with no omissions.}
    \item The key for each JSON object must be the event's \texttt{event\_index}.
\end{enumerate}

\vspace{\baselineskip}

\textbf{Field Specifications:}
For each event, generate a JSON object containing the following fields:
\begin{itemize}
    \item \texttt{"image\_description"}: (If an image is present) \textbf{A detailed description of the screen content, with a strong focus on the user's area of operation and any visual changes.} For tables seen for the first time, their row and column structure must be fully documented here. If no image is present, the value for this field should be an empty string \texttt{""}.
    \item \texttt{"event\_description"}: \textbf{A precise and objective report of the operation.} It must be strictly based on the operational record, clearly stating what action the user performed on which element in which application, including specific parameters (like input text). Back-reference the context to complete all necessary linked information.
\end{itemize}

\vspace{\baselineskip}

\textbf{Output Format Example}

\begin{lstlisting}[language=json]
{
    "0": {
        "image_description": "The screen displays a complete Excel sheet. Visible row headers include: 'Zhang San', 'Li Si', ... Visible column headers include: 'Name'(Col A), 'Age'(Col B), ..., 'Test Subject Fee'(Col I). No specific cell is currently selected.",
        "event_description": "User is viewing a Word window named 'Task Instruction Document' to prepare for the upcoming PPT creation task."
    },
    "2": {
        "image_description": "In the Word document, the user's mouse cursor drags from the end of the text 'The Scientific Basis of a Healthy Lifestyle' backwards, causing this text to be highlighted with a blue background.",
        "event_description": "In the 'Task Instruction Document - Word' window, the user selects the text 'Lecture Title: The Scientific Basis of a Healthy Lifestyle' via a `Drag` operation."
    },
    "4": {
        "image_description": "",
        "event_description": "Following the previous selection, the user executes a `Copy` action, copying the text 'The Scientific Basis of a Healthy Lifestyle' to the system clipboard."
    },
    "8": {
        "image_description": "The screen shows a PowerPoint window. In the slide editing area, after the user right-clicks on an element named 'Title Text Box', a context menu appears. The menu includes 'Paste Options', and the third icon, 'Keep Text Only', is highlighted.",
        "event_description": "The user switches to the '1 - PowerPoint' application and clicks the 'Keep Text Only' button from the 'Paste Options' in the right-click menu, preparing to paste the previously copied content."
    }
}
\end{lstlisting}

\vspace{\baselineskip}

\textbf{Final Checklist}
Before outputting, please perform a self-check:
\begin{itemize}
    \item [ ] Does the number of output events exactly match the number of input events?
    \item [ ] Does each "event\_description" accurately and objectively reflect the operational log?
    \item [ ] For table operations, have "Global Structure Memory" and "Cross-Screenshot Inference" been correctly applied?
    \item [ ] Are sequential operations (like copy-paste) described with proper linkage and information completion?
\end{itemize}
\end{quote}

\section{The Prompt of Cognitive Chain Extractor}
\label{appendix:prompt_cogchain_extractor}

\textit{system}

\begin{quote}
\sloppy
\textbf{Role: Intelligent Analysis Expert for User Computer Behavior}

\vspace{\baselineskip}

\textbf{Core Task}

Your core task is to construct a coherent narrative of the user's cognitive activities \textbf{prior to executing each operation}, based on a series of \texttt{event-image pairs} and historical summaries. This narrative must include quantitative parameters.

\textbf{Primary Directives}:
\begin{enumerate}
    \item \textbf{Completeness and Context}: Generate an analysis for \textbf{every event} in the current batch, based on the `[summary from the previous batch]`.
    \item \textbf{Multimodal Analysis}: \textbf{Synthesize analysis of the image (operational context) and text (operational summary)} to infer the user's cognitive process.
    \item \textbf{Unified Output}: Your final output \textbf{must} be a single, completely formatted JSON object.
\end{enumerate}

\vspace{\baselineskip}

\textbf{Thinking Framework (Core Algorithm - Must Be Strictly Followed)}

\textbf{Step 1: Generate Iterative Summary}

Combine the `[summary from the previous batch]` with a quick preview of all current `event-image pairs` to generate a summary in the `task\_summary` field of the final JSON output.

\textbf{Step 2: Analyze Events One by One}

For each `event-image pair`, strictly adhere to the following algorithm:

\textbf{A. Determine Subtask Status}
\begin{itemize}
    \item \textbf{Compare and Update}: Determine if the user's subtask goal has \textbf{significantly changed}. If it has not, the `current\_subtask` field \textbf{must copy} the content from the previous event. If it has changed, you \textbf{must define} a new subtask goal.
\end{itemize}

\textbf{B. Construct the Preceding Cognitive Chain (Based on the determination in Step A)}
\begin{itemize}
    \item \textbf{Core Paradigm: Cognitive Precedence}: The `cognitive\_chain` you generate \textbf{must be a series of preceding thoughts that led to the occurrence of the currently recorded event}. This means you need to combine the context of the previous event to deduce what the user was thinking and preparing in their mind \textbf{before} executing the \textbf{current} operation.

    \item \textbf{Mandatory Linkage Rule: Triggering and Executing `Orient` and `Verify` (Cannot be violated)}:
    \begin{itemize}
        \item \textbf{Detect}: When you are analyzing an event (e.g., Event N) and \textbf{first determine} in Step A that its `current\_subtask` is \textbf{different} from the previous event (Event N-1), you have triggered this rule.
        \item \textbf{Backtrack and Modify}: You \textbf{must immediately backtrack} to the `cognitive\_chain` you have already generated for Event N-1 and \textbf{append a `Verify` type object to its end}. This is a mandatory modification action.
        \item \textbf{Orient}: After completing the modification for Event N-1, return to the analysis of Event N and \textbf{add an `Orient` type object} to its `cognitive\_chain`.
    \end{itemize}
    
    \item \textbf{Core Distinction: Thinking vs. Executing}:
    \begin{itemize}
        \item \textbf{Exploratory Thinking}: When the user is browsing the interface to make a decision (e.g., scrolling a page to find information), the cognitive chain should primarily contain thinking types like `Find`, `Decide`, `Compute`. The intermediate physical actions (like scrolling) are carriers of thought and \textbf{should not} be modeled as `Execute`.
        \item \textbf{Deterministic Execution}: When the user has a clear goal and begins a series of continuous operations aimed at completing that goal (e.g., a Ctrl+Click+Click+Release multi-select operation), the cognitive chains for these sequential events should primarily be modeled as `Execute`, because the core thinking was completed in preceding steps.
    \end{itemize}
\end{itemize}

\vspace{\baselineskip}

\textbf{Input and Output Format}

\textbf{Input Format}

\textbf{[Summary from the previous batch]}
(...content...)

\textbf{[Events for the current batch]}
(You will receive a series of alternating images and text here)

\vspace{\baselineskip}

\textbf{Output Format}

Your output \textbf{must} be a complete, single JSON object, strictly adhering to the following structure.

\begin{lstlisting}[language=json]
{
  "task_summary": {
    "review": "Review: ...",
    "current": "Current: ...",
    "outlook": "Outlook: ..."
  },
  "event_analysis": [
    {
      "index": "...",
      "reasoning": "Here, based on the image and text, clearly articulate...",
      "details": {
        "event_des": "Original event description",
        "current_subtask": "Clearly describe the subtask goal to which the current operation belongs...",
        "cognitive_chain": [
           {
            "type": "Specific thinking type (must be one of the 11 defined in the toolbox)",
            "content": "Describe in detail the specific manifestation of this thinking type in this scenario...",
            "parameters": {
              "param_name_1": "value (must be the parameter specified for this type in the toolbox)"
            }
          }
        ]
      }
    }
  ]
}
\end{lstlisting}

\vspace{\baselineskip}

\textbf{Cognitive Classification Toolbox (with Mandatory Rules)}

\textbf{Core Task Planning and Orientation}
\begin{enumerate}
    \item \textbf{Orient (Subtask Orientation)}
        \begin{itemize}
            \item \textbf{Definition}: Determine the next (sub)goal and plan the steps to achieve it.
            \item \textbf{Rule}: \textbf{When `current\_subtask` changes, this type must appear in the cognitive chain of the current event. The event with `index: 0` must also contain `Orient`}.
            \item \textbf{Parameter Estimation}:
                 \begin{itemize}
                    \item `steps\_old` (integer): \textbf{The total number of events from the start of the task (index=0) up to the one before the current event, reflecting cumulative load.}
                    \item `steps\_new` (integer): Estimated number of steps for the new subtask that is about to begin.
                \end{itemize}
        \end{itemize}
    \item \textbf{Verify (Verification)}
        \begin{itemize}
            \item \textbf{Definition}: Check the correctness of a manual operation.
            \item \textbf{Rule}: \textbf{When the next event initiates a new `Orient`, this type must be appended to the end of the current event's cognitive chain.}
            \item \textbf{Parameter Estimation}:
                \begin{itemize}
                    \item `m` (integer): The amount of information that needs to be verified.
                \end{itemize}
        \end{itemize}
\end{enumerate}

\vspace{\baselineskip}

\textbf{Cognitive and Physical Operations}
\begin{enumerate}[resume]
    \item \textbf{Find (Visual Localization)}
    \begin{itemize}
        \item \textbf{Definition}: Conduct a visual search in the interface to \textbf{locate} a target element.
        \item \textbf{Rules}:
            \begin{itemize}
                \item If the operation occurs immediately after localization (with no other cognitive activity), there is \textbf{no need} to add an extra `Decide`.
                \item If the element has been located before or its general position is known, `n` \textbf{must} be downgraded to `1`.
                \item \textbf{Example}: Searching for the 'Bold' button in the PowerPoint toolbar, surrounded by similar options like 'Italic' and 'Underline', `n` should be estimated as 3 or higher.
            \end{itemize}
        \item \textbf{Parameter Estimation}:
            \begin{itemize}
                \item `n` (integer): The number of candidate elements.
            \end{itemize}
    \end{itemize}
    
    \item \textbf{Extract (Information Extraction)}
    \begin{itemize}
        \item \textbf{Definition}: \textbf{Read and understand} a piece of text or an image to obtain specific information.
        \item \textbf{Parameter Estimation}:
            \begin{itemize}
                \item `m` (integer): The ratio of the total amount of information on the current page to the amount of information that needs to be extracted.
            \end{itemize}
    \end{itemize}
    
    \item \textbf{Recall (Memory Retrieval)}
    \begin{itemize}
        \item \textbf{Definition}: \textbf{Retrieve} information from memory to guide the current operation.
        \item \textbf{Parameter Estimation}:
            \begin{itemize}
                \item `d` (integer): The number of operational events between the last time this information was stored and the current step.
            \end{itemize}
    \end{itemize}
    
    \item \textbf{Decide (Decision Making)}
    \begin{itemize}
        \item \textbf{Definition}: Evaluate and choose among options \textbf{within} a subtask.
        \item \textbf{Rule}: The `n` and `c` parameters \textbf{must be mutually exclusive}. Use `n` if the decision involves selecting from multiple visible options on the interface; use `c` for strategic or directional implicit decisions.
        \item \textbf{Parameter Estimation}:
            \begin{itemize}
                \item `n` (integer): The number of \textbf{explicit} options (in this case, `c` must be 0).
                \item `c` (float from 0 to 1): An estimate of the size of the \textbf{implicit} decision space (in this case, `n` must be 0).
            \end{itemize}
    \end{itemize}
    
    \item \textbf{Create (Content Generation)}
    \begin{itemize}
        \item \textbf{Definition}: \textbf{Synthesize, infer, and generate new, original content}, not just copy or transcribe.
        \item \textbf{Rule}: This type is used for creative activities requiring mental effort, such as \textbf{devising a summary title, composing the body of an email, or writing a conclusion from scattered data}. Simple keyboard input (like transcribing data) should be modeled as `Recall` -> `Execute`.
        \item \textbf{Parameter Estimation}:
            \begin{itemize}
                \item `m` (integer): The number of independent information points contained in the generated content.
            \end{itemize}
    \end{itemize}
    
    \item \textbf{Compute (Calculation)}
    \begin{itemize}
        \item \textbf{Definition}: Perform mathematical calculations in one's head.
        \item \textbf{Parameter Estimation}:
            \begin{itemize}
                \item `c` (float from 0 to 1): Estimated computational complexity (e.g., comparing two numbers in an image, c$\approx$0.1).
            \end{itemize}
    \end{itemize}
    
    \item \textbf{Execute (Execution)}
    \begin{itemize}
        \item \textbf{Definition}: Translate one or more cognitive steps into a \textbf{final, meaningful physical action}.
        \item \textbf{Rules}:
            \begin{itemize}
                \item Only used to mark the \textbf{end of a thought chain} or a \textbf{step in a deterministic execution sequence}.
                \item An `Execute` \textbf{should not} be created for every scroll or click during an exploratory process.
            \end{itemize}
        \item \textbf{Parameter Estimation}:
            \begin{itemize}
                \item (No parameters)
            \end{itemize}
    \end{itemize}
\end{enumerate}

\vspace{\baselineskip}

\textbf{Output Example (Reinforcing Rule Application)}

\textbf{Input:} 

\textbf{[Summary from the previous batch]}
...The user has completed formatting the main content of the slide.

\textbf{[Events for the current batch]}

\textbf{[Image for Event 48]}

\textbf{[Text for Event 48]} \{ "index": 48, "Event Description": "User clicked the font color button, changing the text color to gray." \}

\textbf{[Image for Event 49]}

\textbf{[Text for Event 49]} \{ "index": 49, "Event Description": "User pressed the Ctrl key in the PowerPoint window." \}

\textbf{[Image for Event 50]}

\textbf{[Text for Event 50]} \{ "index": 50, "Event Description": "User selected the title text box in the PowerPoint window." \}

\vspace{\baselineskip}

\textbf{Output:}
\begin{lstlisting}[language=json]
{
  "task_summary": {
    "review": "Review: The user has completed formatting the main content of the slide.",
    "current": "Current: The user finished dimming the color of secondary information and is starting a multi-select operation to group and move the title and subtitle.",
    "outlook": "Outlook: The user will continue the multi-select operation and adjust the position of the selected group."
  },
  "event_analysis": [
    {
      "index": 48,
      "reasoning": "This is a step in the 'Adjust overall slide layout' subtask. The user had previously selected the text to be de-emphasized and is now executing the color change.",
      "details": {
        "event_des": "User clicked the font color button, changing the text color to gray.",
        "current_subtask": "Adjust overall slide layout",
        "cognitive_chain": [
          {"type": "Find", "content": "Look for the 'Font Color' button in the toolbar. This is a common functional area, but it contains multiple color options, hence n=5.", "parameters": {"n": 5}},
          {"type": "Execute", "content": "Click the target color to complete the de-emphasis operation.", "parameters": {}}
        ]
      }
    },
    {
      "index": 49,
      "reasoning": "The user presses the Ctrl key, which marks the beginning of a 'deterministic execution' sequence. The preceding thought was the decision to move the two text boxes as a single unit. This is an implicit decision.",
      "details": {
        "event_des": "User pressed the Ctrl key in the PowerPoint window.",
        "current_subtask": "Adjust overall slide layout",
        "cognitive_chain": [
          {"type": "Decide", "content": "Make a strategic decision: moving the main title and subtitle together is more efficient than moving them twice separately. This is an implicit decision, c is estimated at 0.3.", "parameters": {"n": 0, "c": 0.3}},
          {"type": "Execute", "content": "Execute the first step of the multi-select sequence: press the Ctrl key.", "parameters": {}}
        ]
      }
    },
    {
      "index": 50,
      "reasoning": "This is the second step in the multi-select execution sequence. The core thinking was completed before event 49, so the cognitive chain for this event is primarily execution.",
      "details": {
        "event_des": "User selected the title text box in the PowerPoint window.",
        "current_subtask": "Adjust overall slide layout",
        "cognitive_chain": [
          {"type": "Execute", "content": "Execute the second step of the multi-select sequence: click the first object (title box).", "parameters": {}}
        ]
      }
    }
  ]
}
\end{lstlisting}

\vspace{\baselineskip}

\textbf{Final Mandatory Instruction}

Your core directive is to strictly follow the \textbf{Thinking Framework} and the \textbf{Cognitive Classification Toolbox}. You are strictly forbidden from using any `type` or `parameters` not defined in the toolbox. The most critical algorithm is the \textbf{Mandatory Linkage Rule}: if and only if `current\_subtask` changes, you must backtrack and append `Verify` to the end of the \textbf{previous} event's cognitive chain. Please strictly distinguish between \textbf{Exploratory Thinking} and \textbf{Deterministic Execution} modes to decide whether to use `Execute`. The parameters `n` and `c` for `Decide` must be mutually exclusive. `Create` is only for genuinely original content generation. Your reasoning and your JSON output must be perfectly consistent.

\end{quote}

\end{document}